\documentclass[aps,pra,twocolumn,superscriptaddress]{revtex4-1}
\usepackage{graphicx}
\usepackage{amssymb}
\usepackage{amsmath}
\usepackage{bm}
\usepackage{xcolor}
 \usepackage{verbatim}
\usepackage{hyperref}
\usepackage{xcolor}
\definecolor{navy}{HTML}{2E3091}
 \hypersetup{ colorlinks=true, urlcolor=navy, citecolor=navy,linkcolor=navy}
 
\begin{document}

\title{Universal Early Coarsening of Quenched Bose Gases}

\author{Junhong Goo}
\affiliation{Department of Physics and Astronomy, Seoul National University, Seoul 08826, Korea}

\author{Yangheon Lee}
\affiliation{Department of Physics and Astronomy, Seoul National University, Seoul 08826, Korea}
\affiliation{Center for Correlated Electron Systems, Institute for Basic Science, Seoul 08826, Korea}

\author{Younghoon Lim}
\affiliation{Department of Physics and Astronomy, Seoul National University, Seoul 08826, Korea}
\affiliation{Center for Correlated Electron Systems, Institute for Basic Science, Seoul 08826, Korea}

\author{Dalmin Bae}
\affiliation{Department of Physics and Astronomy, Seoul National University, Seoul 08826, Korea}
\affiliation{Center for Correlated Electron Systems, Institute for Basic Science, Seoul 08826, Korea}

\author{Tenzin Rabga}
\affiliation{Center for Correlated Electron Systems, Institute for Basic Science, Seoul 08826, Korea}

\author{Y. Shin}
\email{yishin@snu.ac.kr}

\affiliation{Department of Physics and Astronomy, Seoul National University, Seoul 08826, Korea}
\affiliation{Center for Correlated Electron Systems, Institute for Basic Science, Seoul 08826, Korea}
\affiliation{Institute of Applied Physics, Seoul National University, Seoul 08826, Korea}


\begin{abstract}

We investigate the early coarsening dynamics of an atomic Bose gas quenched into a superfluid phase. Using a two-step quench protocol, we effectively control the cooling rates, $r_1$ and $r_2$, during and after passing through the critical region, respectively, and measure the number of quantum vortices spontaneously created in the system. The latter cooling rate $r_2$ regulates the temperature during the condensate growth, consequently controlling the early coarsening dynamics in the defect formation. We find that the defect number shows a scaling behavior with $r_2$ regardless of the initial cooling rate $r_1$, indicating universal coarsening dynamics in the early stage of condensate growth. Our results demonstrate that early coarsening not only reduces the defect density but also affects its scaling with the quench rate, which is beyond the Kibble--Zurek mechanism.

\end{abstract}

\maketitle

When a system crosses a symmetry-breaking phase transition, topological defects can be spontaneously created. This defect formation originates from the causal independence of distant regions and is a generic process in non-equilibrium phase transition dynamics, which gives it broad relevance in condensed matter physics as well as cosmology~\cite{Kibble76,Zurek85,Zurek96,Dziarmaga10}. A practically important problem is the quantitative estimation of the created defect density. However, this is a challenging problem that requires a full description of the complex phase transition dynamics, including the emergence and coarsening of the order parameter and subsequent defect formation and relaxation.

The Kibble--Zurek mechanism (KZM) provides a general framework for defect density estimation~\cite{Zurek85,Zurek96,Dziarmaga10}, where the system's correlation length $\xi$, after passing through the critical region, is assessed for a given quench rate using the system's equilibrium properties near the critical point~\cite{Hohenberg77}, which is assumed to determine the characteristic length scale of the spatial domains of the symmetry-broken phase and consequently, the defect density. The relationship between the defect density and the quench rate was predicted to follow a universal power-law, and has been tested by many experiments~\cite{Hendry94,Bauerle96,Ruutu96,Chuang91,Carmi00,Monaco02,Pyka13, Ulm13, Ejtemaee13,Weiler08, Lamporesi13,Corman14,Navon15,Chomaz15,Sadler06,Ko19,Donadello16,Goo21,Liu21}. Nevertheless, the Kibble-Zurek (KZ) theory inherently lacks the ability to quantitatively predict the average value of defect density because it omits the defect formation dynamics after the freeze-out period in the critical region~\cite{Biroli10,Chandran13,Gagel15,Chesler15}.

In order to produce well-defined topological defects, the order parameter must grow sufficiently, and during the growth period, the spatial fluctuations of the order parameter is inevitably coarsened, affecting the defect formation probability. We refer to this coarsening as {\it early} coarsening, distinguished from the coarsening at later times, where defects decay as the system relaxes to an equilibrium state (Fig.~1). Some theoretical studies have shown that the early coarsening, along with possible coarsening within the critical region~\cite{Biroli10}, simply gives a logarithmic correction to the defect density~\cite{Zurek96,Das12}; but it was recently asserted that it can be sufficient enough to cause defect density saturation for fast enough quenches~\cite{Chesler15}. Such defect density saturation was observed in ultracold atomic gas experiments~\cite{Ko19,Donadello16,Goo21,Liu21}, and furthermore, it was demonstrated that the saturation is not caused by rapid annihilation of defects due to their high density but is possibly associated with the postquench condensate growth dynamics~\cite{Goo21}.

In this paper, we present a direct observation of the early coarsening effect in spontaneous defect formation in a thermally quenched atomic Bose gas. We use a two-step quench protocol, where the sample is successively cooled with two different quench rates so that the temperature change during and after passing through the critical region can be effectively controlled separately. We observe that the mean number of created quantum vortices is reduced as the latter quench rate decreases, and furthermore, the suppression factor is independent of the initial quench rate, indicating the existence of universal early coarsening dynamics in the quenched system. Our results clearly demonstrate the early coarsening effect in spontaneous defect formation and provide a deeper perspective on the universal scaling of the defect density with quench rate, beyond the original KZM.

\begin{figure}[t]
	\includegraphics[width=7.4cm]{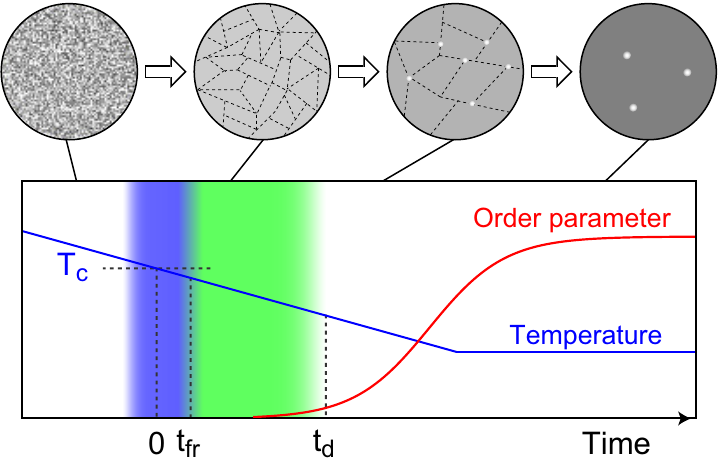}
		\caption{Early coarsening in a quenched Bose gas. As a thermal Bose gas is cooled into a superfluid phase, it experiences an early coarsening period (green), after passing through the critical region (blue) but before the order parameter has grown significantly ($t_\textrm{fr}<t<t_d$), during which the spatial fluctuations of the system are coarsened, reducing the defect creation probability. $T_c$ denotes the critical temperature. In the critical region, the system's dynamics is effectively frozen due to the divergence of its relaxation time at the critical point. In the upper row, the system's quench evolution is illustrated, where the boundaries of spatial domains of the symmetry-broken phase are indicated by dashed lines and quantum vortices are denoted by the white circles.}
\end{figure}

\begin{figure}[t]
	\includegraphics[width=8.5cm]{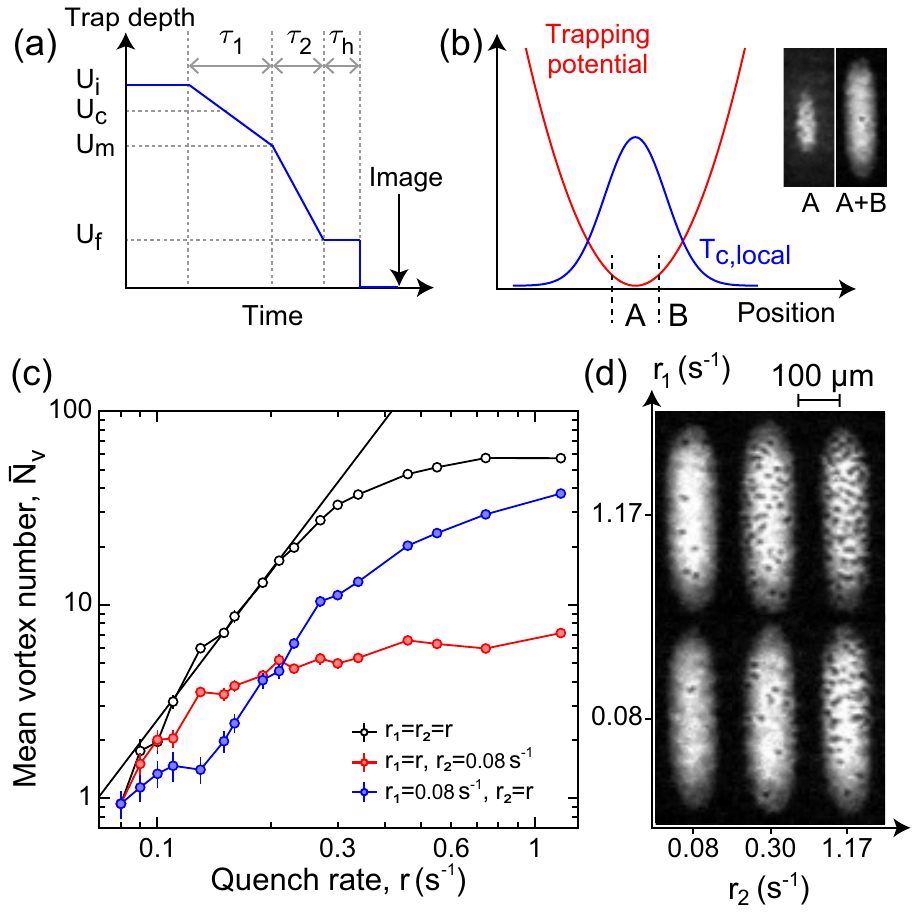}
		\caption{Two-step quench experiment. (a) Schematic of the quench protocol. The trap depth $U$ is linearly lowered from $U_i$ to $U_m$ for time $\tau_1$ with rate $r_1$ and from $U_m$ to $U_f$ for $\tau_2$ with rate $r_2$. $U_c$ denotes the critical trap depth for Bose--Einstein condensation. After a hold time $\tau_h$, a time-of-flight image is taken to detect created vortices. (b) For a trapped sample, the local critical temperature $T_{c,local}$ spatially varies over the sample. The inner (outer) region A (B) undergoes phase transition for $U>U_m (< U_m)$. The inset shows images of samples in equilibrium at $U=U_m$ (left) and $U=U_f$ (right). (c) Mean vortex number $\bar{N}_v$ as a function of quench rate on log--log axes. Open circles denote the data obtained with $r_1=r_2$ and red (blue) solid circles show the data for fixed $r_2(r_1)=0.08~\textrm{s}^{-1}$. The solid line shows a power-law function with exponent $\alpha_\textrm{KZ}=2.6$, fit to the data with $r_1=r_2$ in the scaling regime. Each data point in the blue and red (black) curves was obtained from 30 (20) realizations of the same experiment. The error bars indicate the standard errors of the mean. If the error bars are invisible, they are smaller than the marker size. (d) Representative images of samples for various $r_1$ and $r_2$, displaying quantum vortices by their expanded, density-depleted cores.}
\end{figure}

Our experiment starts by preparing a cold thermal $^{87}$Rb gas in an optical dipole trap (ODT) with a highly oblate and elongated geometry~\cite{Goo21,Lim21}. The initial sample contains $\approx3.3\times10^{7}$ atoms at a temperature of $\approx 480$~nK. Then, the sample is cooled by lowering the ODT depth $U$ from $U_i=1.15U_c$ to $U_m=0.8U_c$ and successively to $U_f=0.27U_c$ in a piece-wise linear manner. Here, $U_c$ is the critical trap depth for Bose--Einstein condensation in an equilibrium sample with $\approx 3.0 \times10^{7}$ atoms. The linear relationship between the trap depth and sample temperature was confirmed~\cite{Goo21}. At the end of the quench, the typical atom number is $\approx 1.2\times10^{7}$, and the sample temperature is $\approx 50$~nK. In equilibrium, the condensate fraction is about 80\%, and the Thomas--Fermi radii are $R_{x,y,z}\approx (65,244,2.8)~\mu\mathrm{m}$. After the quench, a hold time of $\tau_h=1.25~\mathrm{s}$ is applied to facilitate defect formation~\cite{Goo21}, and the created vortices are detected by imaging the sample after a time-of-flight of 40.4~ms.

Figure~2(a) shows a schematic of the two-step quench protocol. The two cooling steps proceed with variable time durations $\tau_1$ and $\tau_2$, giving the quench rates $r_1=\frac{U_i-U_m}{U_c}\times\frac{1}{\tau_1}$ and $r_2=\frac{U_m-U_f}{U_c}\times\frac{1}{\tau_2}$, respectively. In our experiment, $r_{1(2)}$ varies from $r_m$=0.08~s$^{-1}$ to $r_M$=1.17~s$^{-1}$, and the intermediate trap depth $U_m=0.8U_c$ is chosen such that the sample has a negligible condensate fraction ($<$ a few~\%) immediately after the first cooling step~\cite{SM}. This ensures that the condensate growth mostly occurs in the second quench period, and thus we can modulate the early coarsening dynamics with the variable quench rate $r_2$. According to previous $^{87}$Rb experiments~\cite{Navon15}, the unfreeze time $t_{\textrm{fr}}$, which is the time required to pass the critical region (Fig.~1), is estimated to be $\approx 55$~ms ($\ll \tau_1$) for our fastest quench, and it is reasonable to assume that the initial seed structure of ordered-phase domains is implanted at the end of the first quench step. 

In the two-step quench experiment, the early coarsening dynamics can be directly investigated by the dependence of the vortex number $N_v$ on $r_{2}$. However, in the analysis of $N_v$, the density inhomogeneity of the trapped sample must be considered~\cite{delCampo11,delCampo13}. The local critical temperature $T_{c,local}$ varies over the sample, which is higher in the high-density central region than in the low-density outer region [Fig.~2(b)], so the phase transition occurs at different times in different regions of the sample during the quench. For the two-step quench, we can divide the sample into two distinct regions: the central region A, where the local phase transition occurs for $U>U_m$, i.e., the first quench period, while for the outer region B, it occurs during the second quench period with $U<U_m$. Figure~2(b) shows two time-of-flight images of the sample in equilibrium at $U=U_m$ and $U=U_f$, indicating that the area of the region A is $\approx36\%$ of the whole sample area. In view of the local density approximation, we model the total vortex number $N_v$ as a sum of two contributions:
\begin{equation}
\begin{aligned}
 N_v=N_A(r_1,r_2)+N_B(r_2),
\label{eq1}
\end{aligned}
\end{equation}
where $N_A$ and $N_B$ represent the numbers of vortices created in regions A and B respectively. Note that $N_B$ is determined only by $r_2$ because the region B remains thermal with $T>T_{c,local}$ during the first quench period. In the following, we analyze our measurement results based on this model.

In Fig.~2(c), we display the mean vortex number $\bar{N}_v$ as a function of the quench rate for three representative cases: (I) $r_1=r_2$, (II) $r_2=r_m$, and (III) $r_1=r_m$. Setting $r_1=r_2$ corresponds to a typical single-step quench, and as observed in~\cite{Goo21}, $\bar{N}_v$ exhibits power-law scaling for slow quench rates, and saturates for quench rates over 0.3~s$^{-1}$. We obtain the power-law exponent $\alpha_\textrm{KZ}=2.6(1)$ from a fit to the data in the scaling regime, which is determined from a saturation model fit to the $\bar{N}_v$ curve~\cite{SM}. The value of $\alpha_\textrm{KZ}$ is slightly smaller than our previous measurement~\cite{Goo21}, which is due to the refinement of our trap center control during the quench~\cite{SM}.

When $r_2$ is fixed at the slowest value $r_m$ (case II), $\bar{N}_v$ initially follows the single-step quench curve but quickly becomes saturated. Since $N_B(r_m)<N_v(r_m,r_m)\approx 1$, the measured $\bar{N}_v$ can be assumed to mainly reflect $N_A(r_1,r_m)$, the vortex number of region A. In comparison with the single-step quench case, the defect saturation occurs at a lower $r_1$, which is accounted for by the condensate fraction being negligible at the end of the first quench step in the experiment. For the fastest $r_1=r_M$, the maximum vortex number is $\bar{N}_{v,max}<10$, and its ratio to $\bar{N}_{v,max}\approx 60$ in the single-step quench case is almost two times smaller than the area ratio $\eta\approx 0.36$ of region A to the whole sample. This means that $N_A(r_M,r_m)<\eta N_v(r_M,r_M)< N_A(r_M,r_M)$, indicating the early coarsening dynamics. The second inequality results from the fact that the central region has a higher defect formation probability than the outer region.

In case III with $r_1=r_m$, the mean vortex number is also reduced compared to the single-step quench, which is due to the suppression of defect formation in region A. It is notable that the increasing rate of $\bar{N}_v$ becomes faster as $r_2$ increases over $0.13~\mathrm{s}^{-1}$. This is caused by a rapid increase in $N_B$ with increasing $r_2$, and we infer that $N_B>N_A$ for $r_2 > 0.13~\mathrm{s}^{-1}$, i.e., there are more vortices in the outer region than in the central region.

\begin{figure}[t]
	\includegraphics[width=7.6cm]{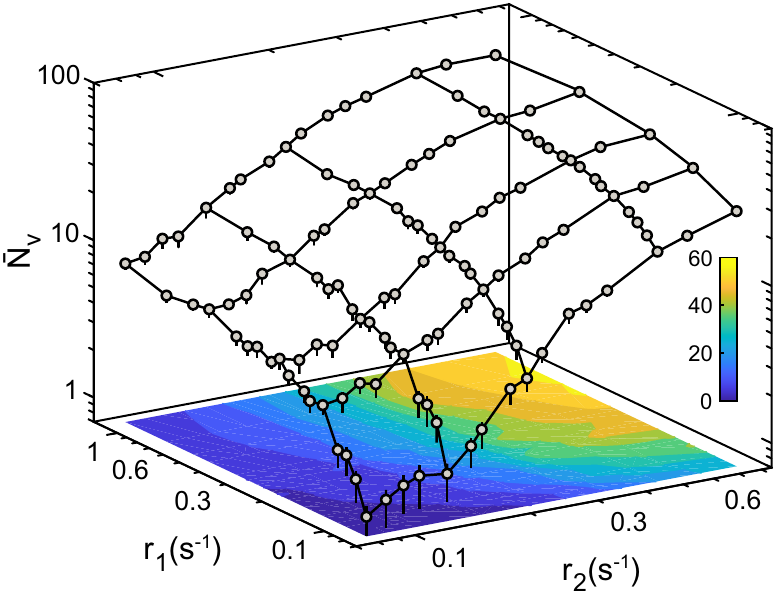}
	\caption{Mean vortex number $\bar{N}_v$ as a function of of $r_1$ and $r_2$ on three-dimensional log axes. Each data point was obtained from 20 realizations of the same experiment, except one with $r_{1(2)}=r_m$ which was obtained from 30 measurements (same data in Fig.~2(c)). The error bars indicate the standard errors of the mean.}
\end{figure}

\begin{figure*}[t]
	\includegraphics[width=17.8cm]{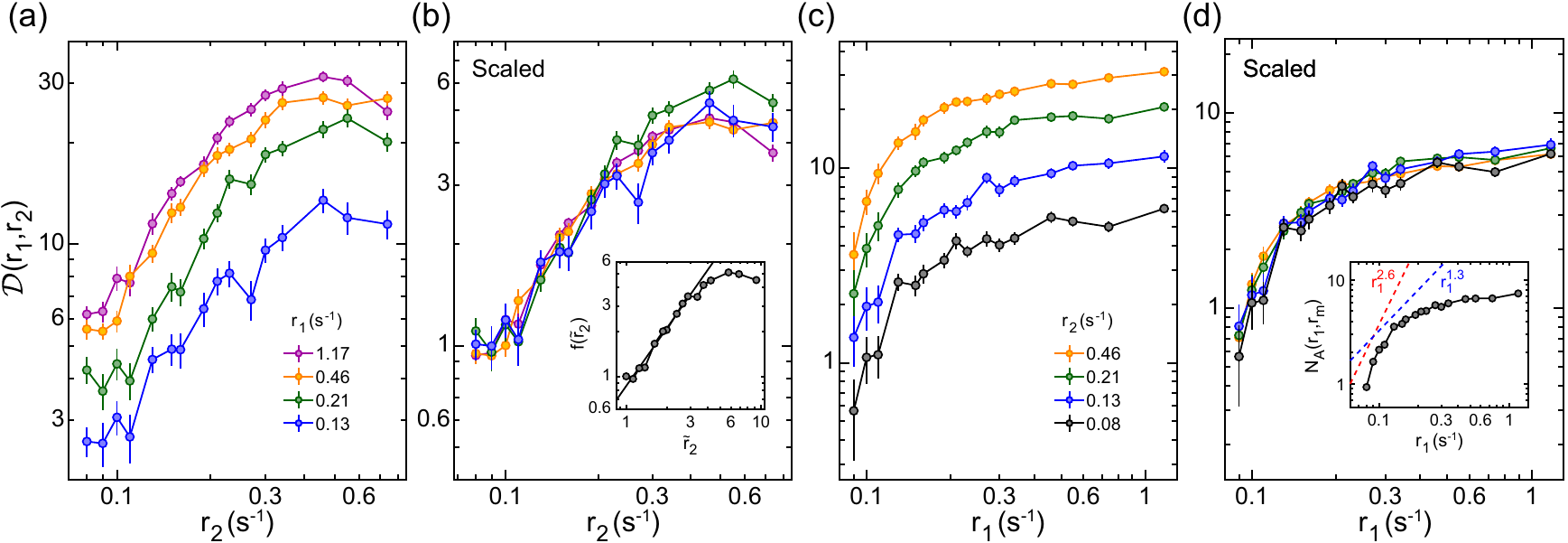}
	\caption{Universal scaling of early coarsening. For a given value of $r_2$, $\mathcal{D}(r_1,r_2)$ denotes the increase in $\bar{N}_v$ as $r_1$ increases from $r_m$ (Eq.~(2)). (a) $\mathcal{D}$ as functions of $r_2$ for four different $r_1$ on log--log axes and (b) the same curves after scaling (see text for details). The inset of (b) presents a universal curve $f(\tilde{r}_2)$ ($\tilde{r}_2=r_2/r_m$) obtained from averaging the data in (b), and the solid line shows a power-law function with exponent $\beta=1.3$, fit to the data in the scaling regime. (c) $\mathcal{D}$ as functions of $r_1$ for four different $r_2$ on log--log axes and (d) the same curves after divided by $f(\tilde{r}_2)$. The inset of (d) presents $N_A(r_1,r_m)$ calculated from the data in (d) (see text for details) and the dashed lines are power-law guide lines. The error bars in (a)--(d) are the standard errors of the measured quantities.}
\end{figure*}

To investigate the early coarsening effect in a systematic manner, we measure $\bar{N}_v$ by scanning over a range of values of $r_1$ and $r_2$. The measurement results are displayed in Fig.~3. As expected from the results as shown in Fig.~2(c), $\bar{N}_v$ monotonically increases with both $r_1$ and $r_2$ and saturates at fast quench rates. To isolate the contribution of $N_A$ to $\bar{N}_v$, we consider the following quantity: 
\begin{equation}
\begin{aligned}
 \mathcal{D} (r_1,&r_2)=\bar{N}_v(r_1,r_2)- \bar{N}_v(r_m,r_2)\\&=N_A(r_1,r_2)-N_A(r_m,r_2).
\end{aligned}
\end{equation}
For a given value of $r_2$, this quantifies the increase in $\bar{N}_v$ as $r_1$ increases from $r_m$. Therefore, the $r_2$ dependence of $\mathcal{D}$ should unambiguously reveal the early coarsening effect in the defect formation process.

In Fig.~4(a), we plot $\mathcal{D}$ as a function of $r_2$ for various $r_1$. We observe that regardless of the value of $r_1$, $\mathcal{D}$ increases over the entire range of $r_2$. This is consistent with the effect of early coarsening, i.e., with a slower second quench at a small $r_2$, the system stays at a high temperature for longer such that the condensate grows slowly, allowing more time for the coarsening of the spatial fluctuations before defects are stably formed, thus reducing $\bar{N}_v$.

Meanwhile, the different plots in Fig.~4(a) show similar profiles for all values of $r_1$. In Fig.~4(b), we replot the four data curves by multiplying each curve by a scaling factor to give the same average value for the four slowest $r_2$. Remarkably, we find that the four scaled curves overlap. This indicates the existence of a universal curve $f(\tilde{r}_2)$ that satisfies the factorization of $\mathcal{D}=f(\tilde{r}_2)\mathcal{D}(r_1,r_m)$ with $\tilde{r}_2=r_2/r_m$. To corroborate this factorization, we plot $\mathcal{D}$ as a function of $r_1$ for four different values of $r_2$ in Fig.~4(c). As can be seen in Fig.~4(d), the four curves overlap after being divided by $f(\tilde{r}_2)$, where the value of $f(\tilde{r}_2)$ was determined by averaging the four curves in Fig.~4(b). The curve overlaps observed in Figs.~4(b) and 4(d) lead us to suggest the following relation
\begin{equation}
\begin{aligned}
N_A(r_1,r_2)=f(\tilde{r}_2)N_A(r_1,r_m),
\end{aligned}
\end{equation}
where $f(\tilde{r}_2)$ represents the suppression factor accounting for the universal coarsening dynamics in the early stage of defect formation. This finding is the main result of this work.

The profile of $f(\tilde{r}_2)$ provides more information on the early coarsening effect. Similar to $\bar{N}_v$ in the single-step quench case, it shows a power-law-like behavior for low $r_2$ and becomes saturated for high $r_2>0.3~\textrm{s}^{-1}$. This is consistent with the postquench condensate growth observed for high $r_2$~\cite{SM}. From a power-law function fit to the data in the scaling regime $r_2\lesssim 0.2$, determined using a saturation model curve as in the single-step quench case~\cite{SM}, we obtain a scaling exponent of $\beta=1.3(2)$ (Fig.~4(b) inset). It is surprising that the measured value is large, approximately half the Kibble--Zurek exponent of $\alpha_\textrm{KZ}=2.6(1)$ in the single-step quench experiment. In previous studies, $\alpha_\textrm{KZ}$ has been associated only with how fast the system passes through the critical region, i.e., the {\it seeding} process, neglecting the early coarsening dynamics. Our observation of large $\beta$ challenges the conventional interpretation of $\alpha_\textrm{KZ}$ based on the KZ theory, demonstrating the significance of the early coarsening dynamics in the defect formation.

An important follow-up question is whether the difference between the two exponents, $\alpha_\textrm{KZ}-\beta\approx 1.3$, can represent the scaling exponent for the KZ seeding process. Currently, we have no clear answer to the question. In the inset of Fig.~4(d), we display $N_A(r_1,r_m)$ as a function of $r_1$, which, as shown in Eq.~(3), is the number of vortices in region A after scaling it by the coarsening factor $f(\tilde{r}_2)$. From Eqs.~(2) and (3), $N_A(r_1,r_m)=\mathcal{D}/f(\tilde{r}_2) + N_A(r_m,r_m)$, and we calculate this by determining $\mathcal{D}/f$ by averaging the data in Fig.~4(d) and assuming $N_A(r_m,r_m)=\bar{N}_v(r_m,r_m)$. A clear power-law behavior is not observed. It should be noted that our analysis is based on the two-region model that ignores possible physics at the regional interface in the inhomogeneous system, such as phase information propagation~\cite{delCampo13,delCampo11}. Further theoretical and experimental investigations are warranted. Similar two-step quench experiments with homogeneous samples~\cite{Navon15,Chomaz15}, in particular, correlating the defect density with the correlation length $\xi$~\cite{Navon15}, would be fruitful to reveal the details of the universal early coarsening dynamics.

In summary, we have investigated the early coarsening in a quenched atomic Bose gas. The two-step quench protocol was introduced to effectively control the coarsening dynamics during the spontaneous defect formation, and the defect number in the central region of the sample was found to be factorizable into functions of each quench rate. Our results demonstrate the existence and characteristics of early coarsening, highlighting the condensate growth dynamics after passing through the critical region. We anticipate that the early coarsening dynamics of the quenched system could be further studied in light of the physics of non-thermal fixed points~\cite{Berges08,Nowak11}, which considers the universal scaling of the spatio-temporal evolution of quantum many-body systems far from equilibrium.

\begin{acknowledgments}
This work was supported by the National Research Foundation of Korea (NRF-2018R1A2B3003373, NRF-2019M3E4A1080400) and the Institute for Basic Science in Korea (IBS-R009-D1). 
\end{acknowledgments}

\section*{Supplemental Material}

\setcounter{equation}{0}
\setcounter{figure}{0}
\setcounter{table}{0}
\makeatletter
\renewcommand{\theequation}{S\arabic{equation}}
\renewcommand{\thefigure}{S\arabic{figure}}
\renewcommand{\thesubsection}{A\arabic{subsection}}

\subsection*{Condensate growth dynamics}

\begin{figure}
	\includegraphics[width=8.0cm]{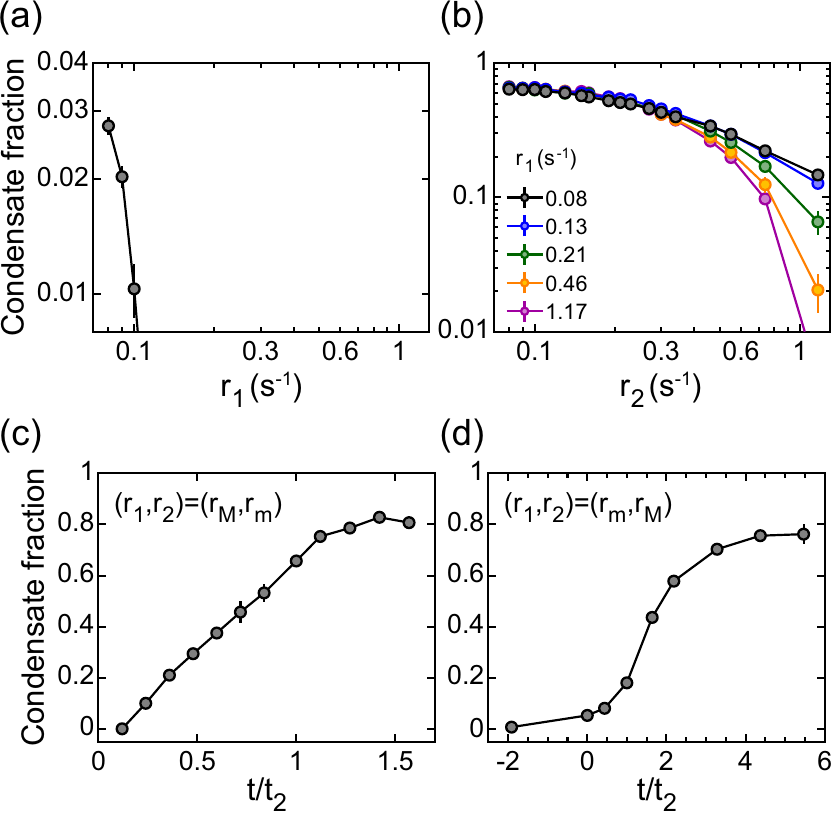}
	\caption{Condensate growth dynamics. (a) Condensate fraction $\eta_{\text{cds}}$ at the end of the first quench step as a function of $r_1$ on log--log axis. (b) $\eta_{\text{cds}}$ at the end of the full two-step quench as a function of $r_2$ for various $r_1$ on log--log axis. (c)(d) Time evolution of $\eta_{\text{cds}}$ in quenching for (c) $(r_1,r_2)=(r_M,r_m)$ and (d) $(r_1,r_2)=(r_m,r_M)$. $t/t_2=0(1)$ corresponds to the start (end) point of the second quench step. Each data point in (a)--(d) comprises three realizations of the same experiment, and its error bar denotes the standard deviation of the measurements.}
\end{figure}

Figure~S1(a) displays the condensate fraction $\eta_\text{cds}$ at the end of the first quench step for various $r_1$. $\eta_\text{cds}$ is negligible ($<1\%$) for most $r_1>0.1~\textrm{s}^{-1}$ and only marginal ($<3\%$) for the slowest $r_1=r_m$. This confirms that most of the condensation growth occurs after the first quench step, and is therefore regulated by the second quench rate $r_2$. In Fig.~S1(b), we display the condensate fraction at the end of the full two-step quench as a function of $r_2$ for various fixed $r_1$. All curves show lagging of the condensate growth for fast $r_2$ and the lagging is more pronounced for faster $r_1$. As shown in Ref.~\cite{Goo21}, the region where the lagging of the condensate growth is severer, corresponds to the region where the defect saturation effect is significant. In Figs.~S1(c) and S1(d), we show the evolution curves of $\eta_{cds}$ for the quenches with $(r_1,r_2)=(r_M,r_m)$ and $(r_1,r_2)=(r_m,r_M)$, respectively. For slow $r_2=r_m$, the condensate grows in a quasi-equilibrium fashion even with the fastest $r_1$, while for fast $r_2=r_M$, the condensate growth is significantly lagging at the end of the second quench even with the slowest $r_1$.

\subsection*{Estimation of $N_B(r_2)$}

Using the coarsening function $f(\tilde{r}_2)$ [Fig.~4(b) inset], the created vortex number for the outer region, $N_B(r_2)$, can be estimated from the measured $\bar{N}_v(r_m,r_2)$ in Fig.~2(c) as 
\begin{equation}
\begin{aligned}
 N_B(r_2)&=\bar{N}_v(r_m,r_2)-N_A(r_m,r_2)\\&=\bar{N}_v(r_m,r_2)-f(\tilde{r}_2)N_A(r_m,r_m)\\&\approx\bar{N}_v(r_m,r_2)-f(\tilde{r}_2)\bar{N}_v(r_m,r_m).
\end{aligned}
\end{equation}
Figure S2 shows the estimated $N_B$ together with the mean vortex number $\bar{N}_v$ in the single-step quench case with $r_1=r_2$. As discussed in the main text, $N_B$ rapidly increases as $r_2$ increases over $0.2~\textrm{s}^{-1}$ and reaches about 30 for the fastest quench rate, which is about 50\% of $\bar{N}_{v,max}\approx 60$. Considering that the area ratio of the outer region to the whole sample is about 64\%, $N_B/\bar{N}_v \sim 0.5$ means that vortices are created somewhat uniformly over the whole sample for the fast quench. $N_A$, the difference between $\bar{N}_v$ and $N_B$ is also plotted in Fig.~S2. For the fastest quench rate, $N_A$ is observed to decrease, which is unphysical, indicating the limitation of our analysis method in the deep saturation regime.

\begin{figure}
	\includegraphics[width=6.3cm]{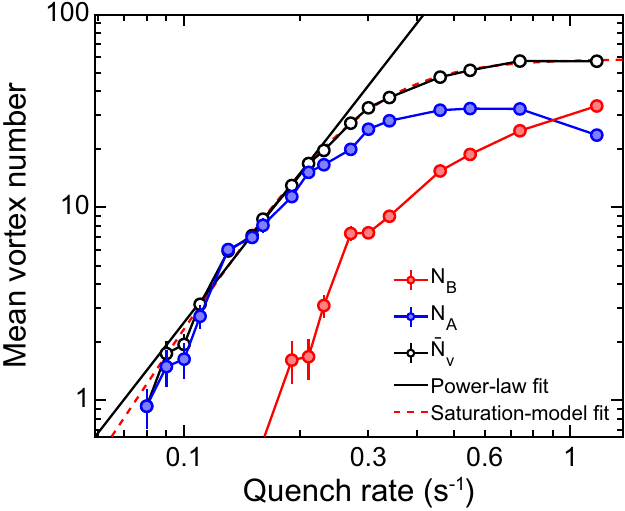}
	\caption{Estimation of $N_A$ and $N_B$ for the single-step quench. Open circles show the measured mean vortex number $\bar{N}_v$ (same in Fig.~2(c)). The solid (dashed) line denotes the power-law (saturation model) function fit to $\bar{N}_v$. The error bars represent the standard errors of the measured quantities.}
\end{figure}

\subsection*{Determination of scaling exponent}
The KZ scaling exponent $\alpha_\textrm{KZ}$ is determined from a power-law function fit to the measurement result of $\bar{N}_v$ (Fig.~S2), but in the fitting, the data points exhibiting saturation behavior should be excluded. To determine the scaling regime for the power-law fit, we use a phenomenological saturation model $N_{\mathrm{S}}(r) =N_{\mathrm{sat}}[1+(r_{\mathrm{sat}}/r)^{\delta\beta_{\mathrm{KZ}}}]^{-1/\delta}$ as described in Ref.~\cite{Goo21}. Here $N_{\mathrm{sat}}$ denotes the saturated vortex number, $r_{\mathrm{sat}}$ the saturation quench rate, $\beta_{\mathrm{KZ}}$ the scaling exponent, and $\delta$ a heuristic parameter for tuning the transition from the scaling behavior to saturation. The fit yields $N_{\mathrm{sat}}=58.8(20)$, $r_{\mathrm{sat}}=0.29(5)$, and $\beta_{\mathrm{KZ}}=3.0(6)$ with $\delta=1.0(4)$. We adopt the scaling regime as $r\leq r_{\mathrm{sat}}/1.5\approx 0.19~\textrm{s}^{-1}$ and obtain $\alpha_\textrm{KZ}=2.6(1)$.

\subsection*{Refinement of trap center control}

In the experiment of the main text, we used a clipped-Gaussian ODT to prepare a large-area sample. A 1064-nm Gaussian laser beam was symmetrically truncated by a horizontal slit and vertically focused through a cylindrical lens to form a highly oblate ODT. As the laser beam is clipped, its focal region is elongated and flattened along the beam propagation axis, resulting in a quartic-like trapping potential~\cite{Lim21}. Because of this weak confinement along the long axis, if the beam axis is not perpendicular to the gravity direction, the gravitational sag can be significant enough to cause a trap center shift with varying ODT power. To compensate for the gravitational sagging, we applied a variable magnetic field gradient and dynamically controlled it to keep the trap center stationary in the horizontal plane during the quench sequence. 

This refined control of the trap center improved the measurement of the scaling exponent $\alpha_\textrm{KZ}$. When the gravitational sag is not properly compensated for, the condensate growth is skewed along the long axis, where the curvature of the trapping potential is slightly higher than that at the original trap center due to the quartic-like profile of the trapping potential. This effect makes the quench dynamics more favorable for the adiabatic generation of the condensate, thus lowering $\bar{N}_v$. We found that the suppression of the vortex number is significant in the slow quench regime, resulting in higher $\alpha_\textrm{KZ}$ without the fine compensation of the trap center shift~\cite{Goo21}.

\end{document}